\documentclass[11pt,a4paper]{iopart}
\usepackage{varioref}
\usepackage[dvips]{graphicx}
\usepackage[latin1]{inputenc}
\usepackage{iopams}

\begin{document}

\vspace{2.5cm}
\begin{center}
{\Large \bf $\Omega^{-}$ and $\bar{\Omega}^{+}$ production in Pb+Pb and p+p collisions at 30, 40 and 158 A$\cdot$GeV}
\end{center}
\vspace{0.3cm}
{\bf Michael Mitrovski}\hspace{0.1cm}{\bf for the NA49 Collaboration}\footnote [6] {Presented at Strange Quark Matter 2003, Atlantic Beach, North Carolina, USA}$^{,}$\footnote [7] {For a full author list of NA49 Collaboration see \cite{A.1}}\\ \\ 
Institut f\"ur Kernphysik, August-Euler-Strasse 6, 60486 Frankfurt, Germany\\ \\
E--mail: Michael.Mitrovski@cern.ch \\

\begin{abstract}
We report preliminary results on $\Omega^{-}$ and $\bar{\Omega}^{+}$ production in central Pb+Pb collisions at 30, 40 and 158 A$\cdot$GeV and p + p  interactions at 158 GeV. The midrapidity $\bar{\Omega}^{+}$/$\Omega^{-}$ ratio is estimated to be 0.45 $\pm$ 0.05 and 0.41 $\pm$ 0.18 for central Pb+Pb collisions at 158 and 40 A$\cdot$GeV, respectively. The corresponding value for 158 GeV p+p interactions is 0.67 $\pm$ 0.62. For central Pb+Pb collisions at 158 A$\cdot$GeV fully corrected distributions are obtained. The inverse slope parameters of the $m_{T}$ spectrum and total yields are $T$($\Omega^{-}$) = 276 $\pm$ 23 MeV, $<$$\Omega^{-}$$>$ = 0.47 $\pm$ 0.07 and $T$($\bar{\Omega}^{+}$) = 285 $\pm$ 39 MeV, $<$$\bar{\Omega}^{+}$$>$ = 0.15 $\pm$ 0.02. 
\end{abstract} \vspace{-0.5cm}

\section{Introduction}

A non-monotonic energy dependence of the $K^{+}$/$\pi^{+}$ ratio with a sharp maximum close to 30 A$\cdot$GeV is observed in central Pb+Pb collisions  \cite{A.1,A.2}. Within a statistical model of the early stage \cite{A.3}, this is interpreted as a sign of the phase transition to a QGP, which causes a sharp change of the energy dependence of the strangeness to entropy ratio. This observation naturally motivates us to study the production of multistrange hyperons ($\Xi$, $\Omega$) as well as the function of the beam energy. Although this is the main motivation for our study of $\Omega$ production in p+p and Pb+Pb collisions, there are two other points of interest. In fact, $\Omega^{-}$ and $\bar{\Omega}^{+}$ production in p+p collisions could give quite detailed information on the dynamics of strangeness production in those collisions. From string-hadronic models, it is expected that the $\bar{\Omega}^{+}$/$\Omega^{-}$ ratio in p+p interactions at SPS energies is higher than 1 whereas it is predicted to be smaller than 1 in a hadron gas model \cite{A.4}. \\
Furthermore it was suggested that the kinematic freeze-out of $\Omega$ takes place directly at hadronization \cite{A.5}. If this is indeed the case, the transverse momentum spectra of the $\Omega$ directly reflect the transverse expansion velocity of a hadronizing QGP \cite{A.6}. \\   
In this report we show preliminary results on $\Omega^{-}$ and $\bar{\Omega}^{+}$ production in central Pb+Pb collisions at 30, 40 and 158 A$\cdot$GeV and compare them to data from p+p  interactions at 158 GeV. We also show fully corrected rapidity and $m_{T}$ spectra of $\Omega^{-}$ and $\bar{\Omega}^{+}$ in central Pb+Pb collisions at 158 A$\cdot$GeV. Integration of the rapidity spectra gives the total yield of $\Omega^{-}$ and  $\bar{\Omega}^{+}$ in full phase space.

\section{The NA49 experiment}

\begin{figure}[h!]
\begin{center}
\includegraphics[scale=0.55]{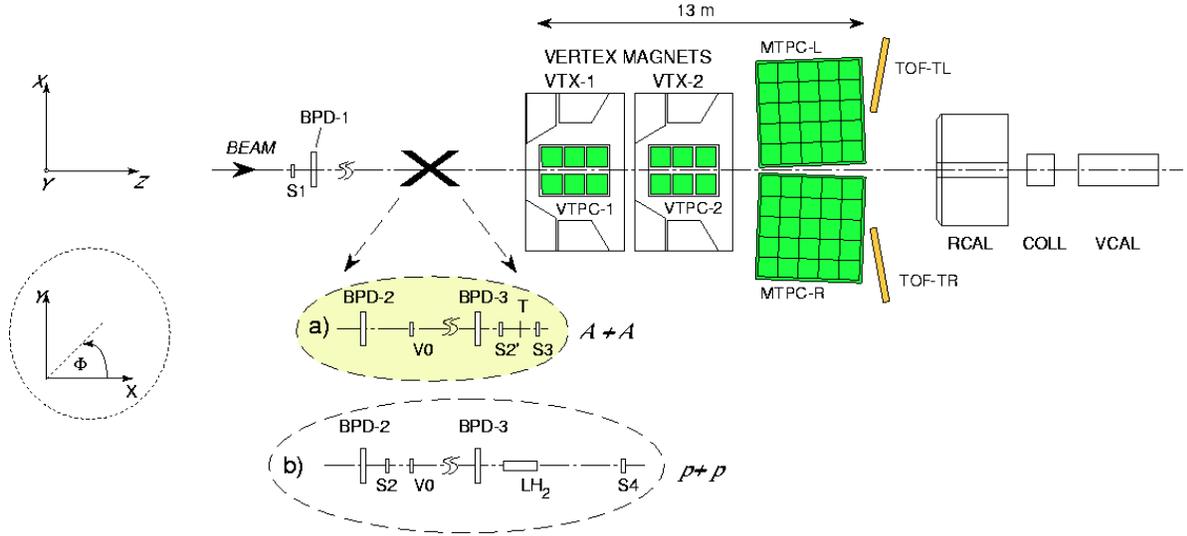} 
\end{center} 
\caption{The NA49 experimental setup. Target configurations used for central Pb+Pb collisions and p+p interactions are shown separately.}  
\label{Experiment}
\end{figure}

The NA49 detector \cite{A.7} (see Fig.\hspace{0.1cm}\ref{Experiment}) is a large acceptance hadron spectrometer at the CERN SPS, consisting of four TPCs. Two of them, the Vertex TPCs (VTPC), are inside a magnetic field for the determination of particle momenta. The ionisation energy loss (dE/dx) measurement in the two Main TPCs (MTPC), which are outside the magnetic field, is used for particle identification. 
 \\
Central collisions were selected by a trigger using information from a downstream calorimeter (VCAL),  which measures the energy of the projectile spectator nucleons. 

\section{Analysis and data sets}

The $\Omega^{-}$ production is analyzed using its decay channel : $\Omega^{-}$ $\rightarrow$ $\Lambda$ $+$ $K^{-}$ (BR=67.8 \%  \cite{A.8}) and $\Lambda$ $\rightarrow$ $\pi^{-}$ $+$ $p$ (BR=63.9 \% \cite{A.8}). For $\bar{\Omega}^{+}$ the charge conjugated channel is used. For the $\Omega^{-}$ ($\bar{\Omega}^{+}$) analysis all $\Lambda$ ($\bar{\Lambda}$) candidates with a reconstructed invariant mass close to the nominal value ($\vert$$\Delta$$M_{\Lambda}$$\vert$ $<$ 0.005 GeV/$c^{2}$) are combined with the $K^{-}$ ($K^{+}$) candidates. In order to identify  the secondary vertex, both are extrapolated back to the target and the point of the closest approach is calculated \cite{A.9}. The resulting combinatorial background is reduced by applying various cuts. The contribution of false $\Lambda$ ($\bar{\Lambda}$) candidates can be reduced by selecting the decay (anti-)proton candidates using their energy loss in the TPCs. The same procedure is used to reduce background in the sample of $K^{-}$ ($K^{+}$) candidates. The reconstructed $\Omega^{-}$ ($\bar{\Omega}^{+}$) candidates have to point to the interaction vertex, while the $K^{-}$ ($K^{+}$) candidate and the $\Lambda$ ($\bar{\Lambda}$) candidate should miss it. Fig. 2 shows the invariant mass distribution of $\Omega^{-}$ ($\bar{\Omega}^{+}$) candidates in central Pb + Pb at 158 A$\cdot$GeV (left), 40 A$\cdot$GeV (middle) and p + p at 158 A$\cdot$GeV (right). Only pairs in the kinematical regions specified in table 1 were used in figures.
\begin{table}[hbt!]
\begin{center}
\begin{tabular}{|c|c|c|c|c|}
\hline
Reaction & Events & $\sigma$/$\sigma_{inel}$ [$\%$] & kinematical region \\[1pt]
\hline
Pb+Pb at 158 A$\cdot$GeV & 3.5 $\cdot$ $10^{6}$  &20  & $\vert$y$\vert$ $<$ 1.0, $p_{t}$ $>$ 0.9 GeV/$c$ \\
Pb+Pb at 40 A$\cdot$GeV & 0.579 $\cdot$ $10^{6}$  &7  & $\vert$y$\vert$ $<$ 0.5, $p_{t}$ $>$ 0.9 GeV/$c$ \\
p+p at 158 GeV & 2.5 $\cdot$ $10^{6}$  & min. bias  & $\vert$y$\vert$ $<$ 0.75 \\ \hline
\end{tabular}
\end{center} 
\caption{Number of analyzed events, percentage of the total inelastic cross section selected by the trigger $\sigma$/$\sigma_{inel}$ and $\Omega^{-}$ ($\bar{\Omega}^{+}$) kinematical region used for the analysis defined by cuts in c.m. rapidity y and transverse momentum $p_{t}$.}
\label{region}
\end{table} \\ \\
A clear signal at the expected mass position ($M_{\Omega}$ = 1.67245 $GeV$$/$$c^{2}$ \cite{A.8}) is seen for all studied reactions. To estimate the number of $\Omega^{-}$ ($\bar{\Omega}^{+}$) decays, the number of entries was integrated in the invariant mass interval [$M_{\Omega}$ $-$ $\Delta$ $M$, $M_{\Omega}$ $+$ $\Delta$ $M$] with $\Delta$ $M$ = 0.010 $GeV$$/$$c^{2}$ and the background contribution was subtracted using a polynominal interpolation. 
\begin{figure}[h]
\begin{center}
\includegraphics[scale=0.6]{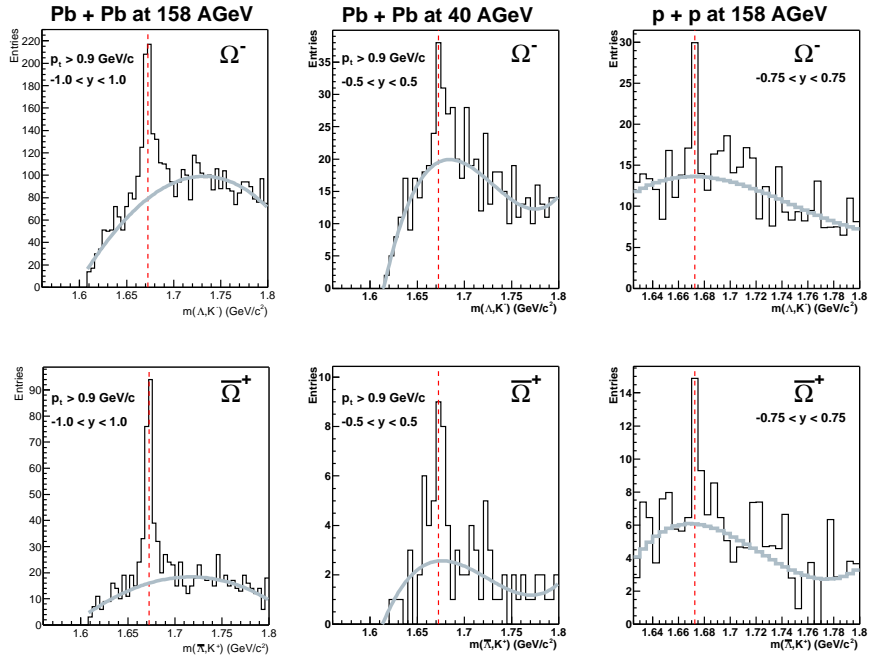} \vspace{-0.3cm}
\end{center}
\caption{The invariant mass distributions of $\Lambda$ $K^{-}$ (upper row) and $\bar{\Lambda}$ $K^{+}$ (lower row) candidate pairs in central Pb + Pb at 158 A$\cdot$GeV (left), 40 A$\cdot$GeV (middle) and p + p at 158 GeV (right). The polynominal parametrisation of the background is indicated by solid lines.}  
\label{Invmass}
\end{figure} \\
For the 158 A$\cdot$GeV Pb+Pb events the raw signals were corrected for the branching ratios, the geometrical acceptance and reconstruction efficiency \cite{A.10}. 

\section{Results}
\subsection{Energy dependence of the $\bar{\Omega}^{+}$/$\Omega^{-}$  ratio}

The procedure presented in the previous section allows us to estimate the uncorrected  $\bar{\Omega}^{+}$/$\Omega^{-}$ ratios in the acceptance given in table 1. The results are given in table 2. 
\begin{table}[hbt!]
\begin{center}
\begin{tabular}{|c|c|}
\hline
Reactions & $\bar{\Omega}^{+}$/$\Omega^{-}$   \\[1pt]
\hline
Pb+Pb at 158 A$\cdot$GeV & 0.45 $\pm$ 0.05  \\
Pb+Pb at 40 A$\cdot$GeV & 0.41 $\pm$ 0.18  \\ 
p+p at 158 GeV  & 0.67 $\pm$ 0.62  \\ \hline
\end{tabular}
\end{center}
\caption{Uncorrected $\bar{\Omega}^{+}$/$\Omega^{-}$ ratios at midrapidity for various studied reactions.} \vspace{-0.3cm}
\label{midratio}
\end{table} \\
The errors are statistical only. In Fig. 3 (left) the NA49 $\bar{\Omega}^{+}$/$\Omega^{-}$ ratios as a function of the collision energy ($\sqrt{s_{NN}}$) are shown and compared with results of WA97/NA57 \cite{A.11,A.12} and STAR \cite{A.13}. The NA49 and WA97/NA57 results measured at the same energies are consistent. The data show a clear increase of the $\bar{\Omega}^{+}$/$\Omega^{-}$ ratio from a value of about 0.5 at SPS energies to about 1 at RHIC energies.  The ratio measured for p+p interactions is similar to the ratio for central Pb+Pb collisions. \vspace{-0.3cm}
\begin{figure}[h!]
\begin{center}
\includegraphics[scale=0.38]{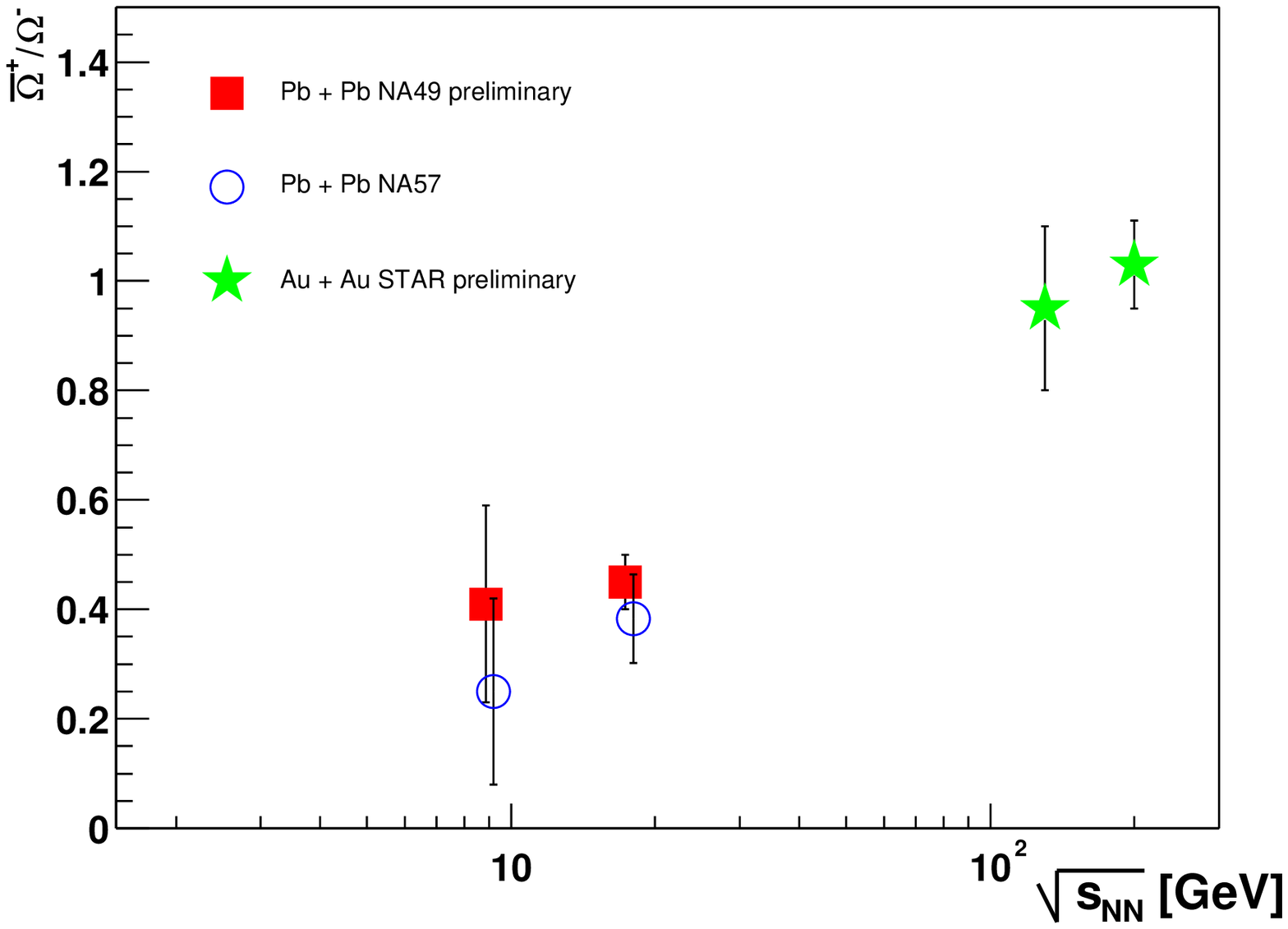}
\includegraphics[scale=0.38]{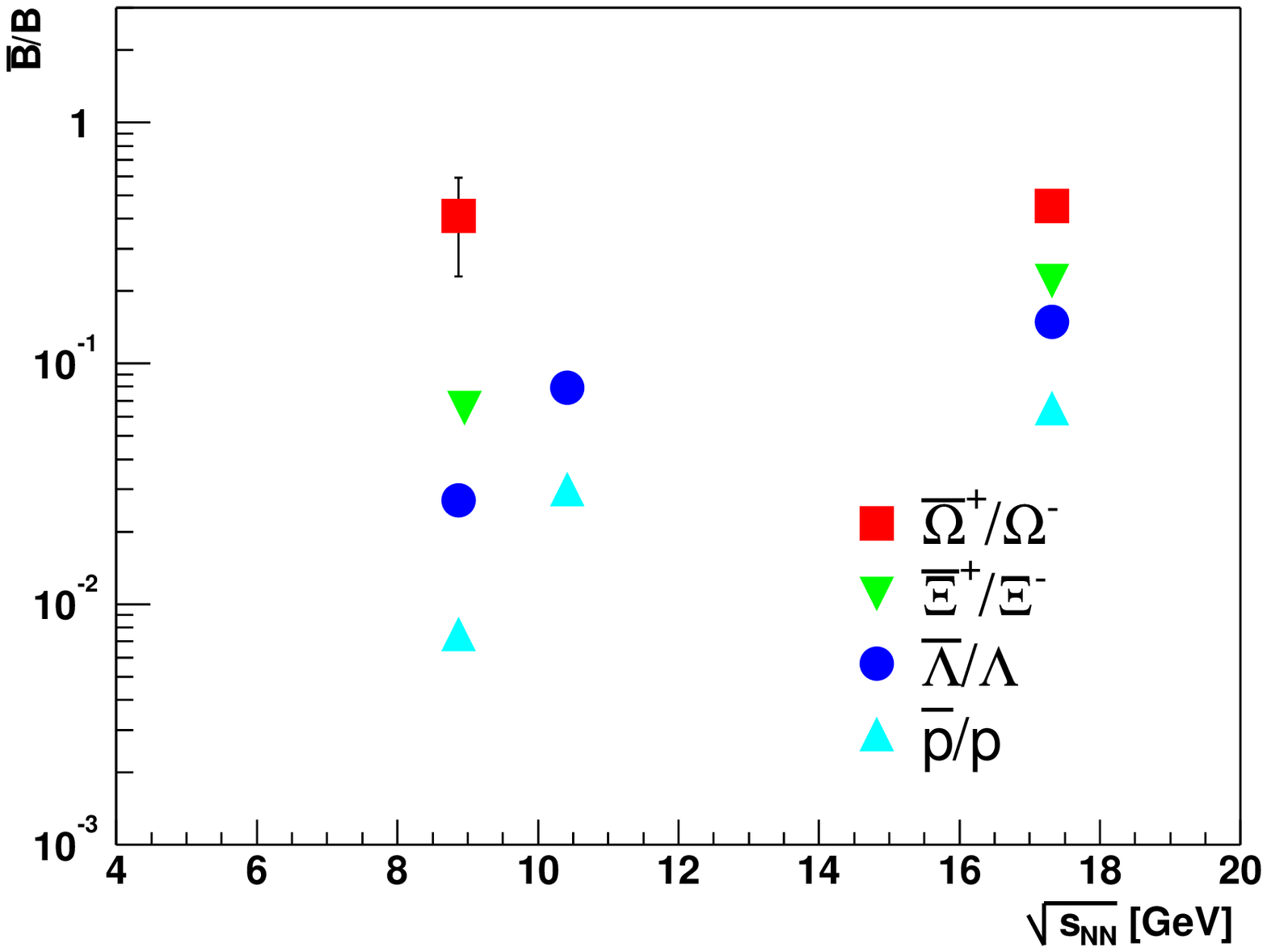}
\end{center}
\caption{\hspace{-0.2cm} The $\bar{\Omega}^{+}$/$\Omega^{-}$ ratio at SPS-RHIC energy range (left) and the antibaryon/baryon ratio ($\bar{B}$$/$$B$) at SPS energy range (right) measured by NA49.}  \vspace{-0.3cm}
\label{Baryon}
\end{figure} \\
In Fig. 3 (right) the antibaryon/baryon ratios are shown as a function of the beam energy in the SPS energy domain. The energy dependence of $\bar{B}$$/$$B$ ratios gets weaker with increasing strangeness content. 

\subsection{$\Omega^{-}$ ($\bar{\Omega}^{+}$) in Pb+Pb at 158 A$\cdot$GeV}

At 158 A$\cdot$GeV, a high statistics data sample of central Pb+Pb collisions is available, which allows us to obtain fully corrected spectra of $\Omega^{-}$ and $\bar{\Omega^{+}}$. The transverse mass $m_{t}$ $=$ $\sqrt{p_{t}^{2} + m_{\Omega}^{2}}$ spectra are shown in Fig. 4. They are fitted by the exponential function \vspace{0.2cm}
\begin{equation}
\frac{1}{m_{t}} \frac{d^{2}N}{dm_{t}dy} = C \cdot e^{- m_{t}/T}  
\end{equation} \vspace{0.3cm}
where the fit parameters are a normalization factor \textit{C} and the inverse slope parameter \textit{T}. The first data point ($m_{t} - m_{0}$ $<$ 0.25 GeV), was excluded from the fit. The slope parameter is similar for $\Omega^{-}$ and $\bar{\Omega}^{+}$ : $T$($\Omega^{-}$) = 276 $\pm$ 23 MeV and $T$($\bar{\Omega}^{+}$) = 285 $\pm$ 39 MeV \cite{A.10}. Our values agree with those measured by the  WA97 collaboration \cite{A.14}. 
\begin{figure}[h!]
\begin{center}
\includegraphics[scale=0.5]{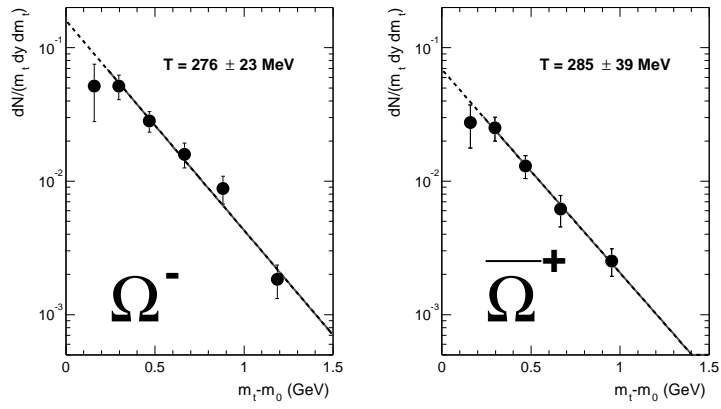}  \vspace{-0.3cm}
\end{center}
\caption{The transverse mass spectra (NA49 preliminary) of $\Omega^{-}$ (left) and $\bar{\Omega}^{+}$ (right) in central Pb + Pb at 158 A$\cdot$GeV, measured in the rapidity range 1.9 $<$ y $<$ 3.9. The lines show fits with an exponential.} 
\end{figure} \\ \\
The large acceptance of the NA49 experiment allows us to measure the $\Omega^{-}$ ($\bar{\Omega}^{+}$) spectra in a large rapidity interval. Fig. 5 shows the rapidity distributions for $\Omega^{-}$ (left) and  $\bar{\Omega}^{+}$ (right) extrapolated to $p_{t}$ = 0 using the exponential shown in Fig. 4. \vspace{-0.5cm}
\begin{figure}[h]
\begin{center}
\includegraphics[scale=0.6]{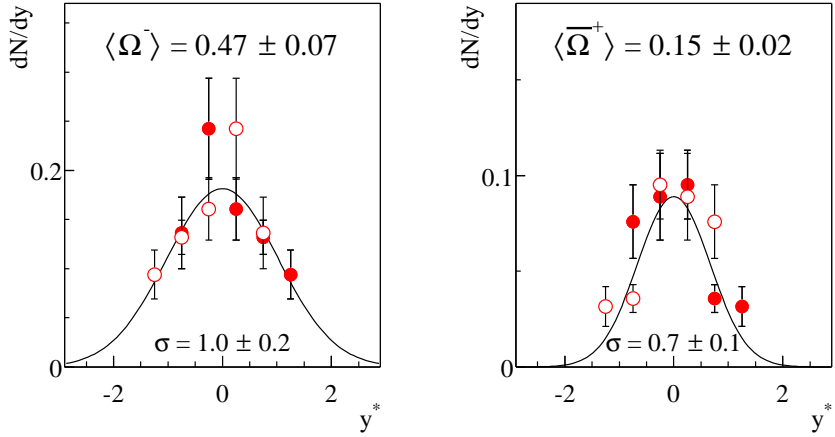} \vspace{-0.8cm}
\end{center} 
\caption{The rapidity spectra (NA49 preliminary) of $\Omega^{-}$ (left) and $\bar{\Omega}^{+}$ (right) in central Pb + Pb at 158 A$\cdot$GeV. The full symbols are the measured points and the open points shows their reflection with respect to midrapidity.} 
\end{figure} \\
Both spectra were fitted by a Gaussian. The width of the $\Omega^{-}$ distribution ($\sigma$($\Omega^{-}$) = 1.0 $\pm$ 0.2) seems to be larger than the one of the $\bar{\Omega}^{+}$ ($\sigma$($\Omega^{-}$)  = 0.7 $\pm$ 0.1). Mean multiplicities in full phase-space were estimated as integrals over measured points corrected for the missing rapidity coverage using the Gaussian parametrisations. The resulting yields are $<$$\Omega^{-}$$>$ = 0.47 $\pm$ 0.07 and $<$$\bar{\Omega}^{+}$$>$ = 0.15 $\pm$ 0.02, where the errors are statistical only. \vspace{-0.2cm}

\section{Conclusions and outlook} \vspace{-0.1cm}

Preliminary results on $\Omega^{-}$ and $\bar{\Omega}^{+}$ production in Pb+Pb collisions and p+p interactions at CERN SPS energies were presented. The $\bar{\Omega}^{+}$/$\Omega^{-}$ ratios increase with energy from about 0.5 at SPS to about 1 at RHIC energies for both central Pb+Pb (Au+Au) and p+p interactions.  \\
The energy dependence of the antibaryon/baryon ratio is weaker for particles with a higher strangeness content. Fully corrected spectra of $\Omega^{-}$ and $\bar{\Omega}^{+}$ were presented at 158 A$\cdot$GeV. \vspace{-0.4cm}
\begin{figure}[h!]     
\begin{center}
\includegraphics[scale=0.4]{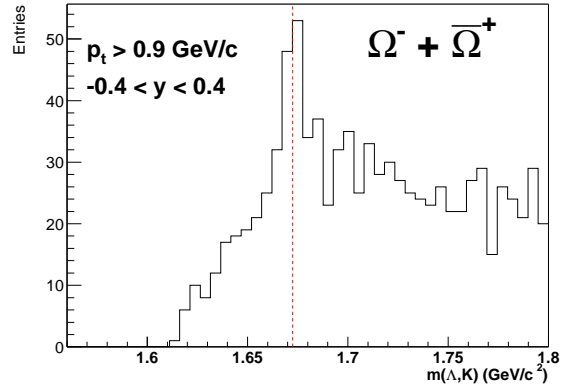} \vspace{-0.3cm}
\end{center} 
\caption{The invariant mass distribution of $\Lambda$ $K^{-}$  $+$ $\bar{\Lambda}$ $K^{+}$ candidate pairs in central Pb + Pb at 30 A$\cdot$GeV.}
\end{figure}\\
In 2002, Pb+Pb collisions at 20 A$\cdot$GeV and 30 A$\cdot$GeV were taken in the interesting range between top AGS and low SPS energies. Fig. 6 shows the first $\Omega^{-}$ + $\bar{\Omega}^{+}$ invariant mass spectrum at 30 A$\cdot$GeV. Analysis of the data at 20, 30 and 80 A$\cdot$GeV is planned for the near future. \vspace{-0.3cm}

\section*{Acknowledgments} \vspace{-0.3cm}
This work was supported by the Director, Office of Energy Research, 
Division of Nuclear Physics of the Office of High Energy and Nuclear Physics 
of the US Department of Energy (DE-ACO3-76SFOOO98 and DE-FG02-91ER40609), 
the US National Science Foundation, 
the Bundesministerium f\"ur Bildung und Forschung, Germany, 
the Alexander von Humboldt Foundation, 
the UK Engineering and Physical Sciences Research Council, 
the Polish State Committee for Scientific Research (2 P03B 130 23, SPB/CERN/P-03/Dz 446/2002-2004, 2 P03B 02418, 2 P03B 04123), 
the Hungarian Scientific Research Foundation (T032648, T14920 and T32293),
Hungarian National Science Foundation, OTKA, (F034707),
the EC Marie Curie Foundation,
and the Polish-German Foundation. 

\end{document}